\documentclass[acmsmall]{acmart}

\usepackage{bm}
\usepackage{url}
\usepackage{hyperref}
\usepackage{tikz}
\usepackage{amsmath}
\usepackage{dsfont}
\usepackage{textcomp}
\usepackage{upgreek}
\usepackage{footnote}
\usepackage{verbatim}
\usepackage{mathrsfs} 
\usepackage{graphicx}
\usepackage{paralist}
\usepackage{multirow}
\usepackage{multicol}
\usepackage{slashbox}
\usepackage{wrapfig}
\usepackage{listings, lstautogobble}
\usepackage{algorithm}
\usepackage{algorithmicx}
\usepackage{algpseudocode}
\usepackage[skip=0cm,list=true,labelfont=it]{subcaption}
\algrenewcommand\textproc{} 
\usepackage{amsmath, times}
\usepackage{cleveref}
\hypersetup{%
	colorlinks=false,
	linkbordercolor=red,
	pdfborderstyle={/S/U/W 1}
}

{}

\AtBeginDocument{%
  }

\setcopyright{acmcopyright}
\copyrightyear{20XX}
\acmYear{20XX}
\acmDOI{XXXXXXX.XXXXXXX}

\acmJournal{JACM}
\acmVolume{00}
\acmNumber{0}
\acmArticle{000}
\acmMonth{0}

\begin{document}
\title[LevelSetPy]{LevelSetPy: A GPU-Accelerated Package for Hyperbolic
	Hamilton-Jacobi Partial Differential Equations}

\author{Lekan Molu}
\email{lekanmolu@microsoft.com}
\orcid{0000-0003-3716-3543}
\affiliation{%
  \institution{Microsoft Research}
  \streetaddress{300 Lafayette St}
  \city{New York}
  \country{USA}
  \postcode{10012}
}

\renewcommand{\shortauthors}{Molu, Lekan}


\received{30 April 2024}

\begin{abstract}
	This article introduces a software package release for geometrically reasoning about the \textit{safety} desiderata of (complex) dynamical systems via level set methods. In emphasis, safety is analyzed with Hamilton-Jacobi equations. In scope, we provide implementations of numerical algorithms for the resolution of Hamilton-Jacobi-Isaacs equations: the spatial derivatives of the associated value function via upwinding, the Hamiltonian via Lax-Friedrichs schemes, and the integration of the Hamilton-Jacobi equation altogether via total variation diminishing Runge-Kutta schemes. Since computational speed and interoperability with other modern scientific computing libraries (typically written in the Python language) is of essence, we capitalize on modern computational frameworks such as \texttt{CUPY} and \texttt{NUMPY} and move heavy computations to GPU devices to aid parallelization and improve bring-up time in safety analysis. We hope that this package can aid users to quickly iterate on ideas and evaluate all possible safety desiderata of a system via geometrical simulation in modern engineering problems. 
\end{abstract}
\maketitle

\definecolor{light-blue}{rgb}{0.2,0.2,0.8}
\definecolor{light-green}{rgb}{0.20,0.49,.85}
\definecolor{purple}{rgb}{0.70,0.69,.2}

\newcommand{\lb}[1]{\textcolor{light-blue}{#1}}
\newcommand{\rev}[1]{\textcolor{red}{#1}}

\renewcommand{\figureautorefname}{Fig.}
\renewcommand{\sectionautorefname}{$\S$}
\renewcommand{\equationautorefname}{equation}
\renewcommand{\subsectionautorefname}{$\S$}
\renewcommand{\chapterautorefname}{Chapter}

\newcommand{\cmt}[1]{{\footnotesize\textcolor{red}{#1}}}
\newcommand{\todo}[1]{\textcolor{cyan}{TO-DO: #1}}
\newcommand{\review}[1]{\noindent\textcolor{red}{$\rightarrow$ #1}}
\newcounter{mnote}
\newcommand{\marginote}[1]{\addtocounter{mnote}{1}\marginpar{\themnote. \scriptsize #1}}
\setcounter{mnote}{0}
\newcommand{\ie}{i.e.\ }
\newcommand{\eg}{e.g.\ }
\newcommand{\cf}{cf.\ }
\newcommand{\yes}{\checkmark}
\newcommand{\no}{\ding{55}}

\newcommand{\flabel}[1]{\label{fig:#1}}
\newcommand{\seclabel}[1]{\label{sec:#1}}
\newcommand{\tlabel}[1]{\label{tab:#1}}
\newcommand{\elabel}[1]{\label{eq:#1}}
\newcommand{\alabel}[1]{\label{alg:#1}}

\newcommand{\bull}[1]{$\bullet$ #1}
\newcommand{\argmax}{\text{argmax}}
\newcommand{\argmin}{\text{argmin}}
\newcommand{\mc}[1]{\mathcal{#1}}
\newcommand{\bb}[1]{\mathbb{#1}}

\newcommand{\shmargin}[2]{{\color{magenta}#1}\marginpar{\color{magenta}\raggedright\footnotesize #2}}
\newcommand{\shnote}[1]%
{\textcolor{magenta}{SH: #1}}
\newcommand{\lmnote}[1]%
{\textcolor{orange}{LM: #1}}

\def\tidx{t}
\newcommand{\Note}[1]{}
\renewcommand{\Note}[1]{\hl{[#1]}}  

\def\kau{\mc{K}}
\def\particle{\bm{x}}
\def\materialresponse{\bm{G}}
\def\orthoggroup{{\textit{SO}}(3)}
\def\liegroup{{\textit{SE}}(3)}
\def\liealgebra{\mathfrak{se}(3)}
\def\identity{\bm{I}}
\newcommand{\trace}[1]{\textbf{tr}(#1)}

\def\flock{F}
\def\rot{{R}}
\def\rthree{\bb{R}^3}
\def\reline{\bb{R}}
\def\targetset{\mathcal{L}}
\def\traj{\xi}
\def\ren{\bb{R}^n}
\def\skew{S}
\def\state{\bm{x}}
\def\statex{x}
\def\statey{y}
\def\statez{z}
\def\hot{h.o.t.\ }
\def\lhs{l.h.s.\ }
\def\rhs{r.h.s.\ }
\def\identity{I}
\def\costdiff{\mathbf{\tilde{V}}}
\def\pursuer{\bm{P}}
\def\evader{\bm{E}}
\def\gain{\bm{k}}
\def\control{\bm{u}}
\def\disturb{\bm{w}}
\def\switchcurve{\bm{\gamma}}
\def\valuefunc{\bm{v}}
\def\valueterm{\bm{g}}
\def\lpspace{L^2({\mc{S}}; \mc{F})}
\def\lpdual{L^2({\mc{S}}; \breve{\mc{F}})}
\def\valuetensor{\mathds{V}}
\def\wtensor{\mathds{W}}
\def\valuecore{\mathds{V}^c}
\def\hamfunc{\bm{H}}
\def\hamtensor{\mathds{H}}
\def\uppervalue{\bm{V}^+}
\def\lowervalue{\bm{v}}
\def\upperham{\bm{H}^+}
\def\lowerham{\bm{H}}
\def\hilbertparam{\bm{\phi}}
\def\hilbertcoeff{\bm{\psi}}
\def\hilbertparamspace{\bm{\Phi}}
\def\hilbertcoeffspace{\bm{\Psi}}
\def\hilbertspace{\mathcal{F}}
\def\hilbertdual{\mathcal{S}}
\def\reducedbasis{\Xi_r}
\def\basis{\mathbf{e}}
\def\openset{\Omega}
\def\spatialdomain{\Omega}
\def\timeinterval{I}
\def\liederi{L}
\def\grid{\textbf{g}}
\def\pde{PDE\,}
\def\payoff{\bm{\phi}}
\def\Payoff{\bm{\Phi}}

\def\group{\mc{C}}
\def\subgroup{\mc{S}}

\newcommand{\cmark}{\ding{51}}%
\newcommand{\xmark}{\ding{55}}%

\newcommand{\win}{\emoji{window}}
\newcommand{\mac}{\emoji{apple}}
\newcommand{\linux}{\emoji{penguin}}
\newcommand{\opensource}{\emoji{magnifying-glass-tilted-right}}
\newcommand{\commercial}{\emoji{money-bag}}

\definecolor{codegreen}{rgb}{0,0.6,0}
\definecolor{codelblue}{rgb}{0,0.0,0.7}
\definecolor{codegray}{rgb}{0.5,0.5,0.5}
\definecolor{codepurple}{rgb}{0.58,0,0.82}
\definecolor{backcolour}{rgb}{0.95,0.95,0.92}

\newcommand*\greencheck{\textcolor{codegreen}{\ding{52}}}
\newcommand*\redcross{\textcolor{red}{\ding{55}}}

\lstdefinestyle{mystyle}{
	backgroundcolor=\color{backcolour},   
	commentstyle=\color{codegreen},
	keywordstyle=\color{magenta},
	numberstyle=\tiny\color{codegray},
	stringstyle=\color{codepurple},
	basicstyle=\ttfamily\footnotesize,
	breakatwhitespace=false,         
	breaklines=true,                 
	captionpos=b,                    
	keepspaces=true,
	frame=single,
	numbers=none,                    
	numbersep=5pt,                  
	showspaces=false,                
	showstringspaces=false,
	showtabs=false,                  
	tabsize=2
}

\lstdefinelanguage{CheckUrdf}
{
	morekeywords={check_urdf},
}

\definecolor{orgred}{rgb}{0.8078,0.4471,0.2314}
\definecolor{darkgreen}{rgb}{0.4157,0.6,0.333}
\definecolor{darkblue}{rgb}{0.0,0.0,0.6}
\definecolor{cyan}{rgb}{0.0,0.6,0.6}
\definecolor{light-gray}{gray}{0.80}
\definecolor{brown}{rgb}{48,19,0}

\lstdefinestyle{xmlStyle}{
	basicstyle=\ttfamily\scriptsize,
	columns=fullflexible,
	showstringspaces=false,
	commentstyle=\color{darkblue},
	morecomment=[l]{//},
	numbers=left,                    
	numbersep=10pt, 
	numberstyle=\tiny,
	captionpos=b,
}
\lstset{style=xmlStyle,escapechar=|}

\lstdefinestyle{pythonStyle}{
	basicstyle=\ttfamily\scriptsize,
	columns=fullflexible,
	showstringspaces=false,
	commentstyle=\color{darkblue},
	morecomment=[l]{//},
	numbers=left,                    
	numbersep=10pt, 
	numberstyle=\tiny,
	captionpos=b,	
	autogobble=true
}
\lstset{style=xmlStyle,escapechar=|}

\lstdefinelanguage{XML}
{
	morestring=[b]",
	morestring=[s]{>}{<},
	morecomment=[s]{<?}{?>},
	stringstyle=\color{black},
	identifierstyle=\color{darkblue},
	keywordstyle=\color{cyan},
	morekeywords={xmlns,type,name,xyz,link,size,radius,length},
}

\definecolor{orange}{rgb}{1,0,0}
\definecolor{grey}{rgb}{0.5,0.5,0.5}
\definecolor{pythoncolor}{rgb}{0.96,0.88,0.76}
\definecolor{joint1color}{rgb}{0.96,0.93,0.89}
\definecolor{others}{rgb}{0.76,0.74,0.82}
\definecolor{salmon}{rgb}{.83,.53,0.53}
\newcommand{\coloropacity}{!65}%
\newcommand{\Hilight}[1]{\makebox[0pt][l]{\color{#1}\rule[-2pt]{0.95\columnwidth}{8pt}}} 

\section{Introduction}

With the growing complexity of digital, electronic, and cyberphysical interfaces in the modern systems that we develop, the need to guarantee performance as envisioned by the systems architect in the face of uncertainty has become ever more timely. To deploy modern systems in the wild, modern software must be able to process generated data at multiple levels of abstraction within reasonable time yet guarantee consistency in system performance \textit{in spite} of the dangers that may evolve if nominally envisioned system performance falter. To address this concern, we consider software architectures for the numerical analysis of the safety assurance (ascertaining the freedom of a system from harm). 
In scope, the computation of safety sets is limited to those numerically obtained via levelset methods~\cite{SethianLSBook} derived from Hamilton-Jacobi (HJ) partial differential equations (PDEs)~\cite{CrandallMethodFracSteps, Evans1984} .

Level sets~\cite{SethianLSBook} are an elegant tool for numerically analyzing the safety of a dynamical system in a reachability context~\cite{LygerosReachability}. They entail discretizing the HJ PDE and employing Lax-Friedrichs Hamiltonian integration schemes~\cite{CrandallLaxFriedrichs}, and the methods of characteristics and lines~\cite{LevelSetsBook} to advance the numerical solution to the HJ equation on a grid along based on a total variation diminishing Runge-Kutta (TVD-RK)~\cite{ShuOsherEfficientENO} scheme. In practice, this is done by constructing an implicit surface representation for a problem's value function and employing \textit{upwinding} to advance the evolution of the dynamical system across time. As can be imagined, updating the values at different nodal points of the state space as integration advances has an exponential computational cost. Most of the existing solvers solve this integration process on single-threaded CPUs, usually with non-optimized routine loops. This is debilitating for many practical problems which tend to be in higher dimensions and whose safety reasoning requires fast computation times. 

The foremost open-source verification software for engineering applications based on HJ equations and levelset methods~ is the CPU-based \texttt{MATLAB}\textregistered\, levelsets toolbox~\cite{MitchellLSToolbox} first developed in 2005. Since its development, there has been vast improvements in hardware and software compute architectures that leverage graphical processing units (GPU) and accelerated linear algebra packages for numerical analysis.  It therefore seems reasonable to provide a means of accelerating the computation of HJ PDE solutions to the end of resolving the safety of many cyberphysical systems in a faster manner. 

In this article, we will describe our efforts in the past three years on designing a \texttt{Numpy}\textregistered\, and GPU-accelerated \texttt{CuPy}~\cite{CUPY} software packages for numerically resolving generalized  discontinuous solutions to time-dependent HJ hyperbolic PDEs that arise in many physical and natural problem contexts. 
\texttt{Numpy} and \texttt{Cupy} are both object-oriented  array programming packages written in the Python language that enable compact and expressive structures for retrieving, modifying, and executing computations on data arranged as vectors, matrices, or higher-order arrays in a computer's memory. Accompanying this package are implicit calculus operations on dynamic codimension-one interfaces embedded on  surfaces in $\mathbb{R}^n$, and numerical discretization schemes for hyperbolic partial differential equations. Furthermore, we describe explicit integration schemes including Lax-Friedrichs, Courant-Friedrichs-Lewy (CFL) constrained  TVD-RK schemes for  HJ equations. Finally, extensions to reachability analyses for continuous and hybrid systems, formulated as optimal control or game theory problems using viscosity solutions to HJ \pde's is described in an example. While our emphasis is on the resolution of safe sets in a reachability context for verification settings, the applications of the software package herewith presented extend beyond control engineering applications.

\subsection{Background and basic notation}
Our chief interest is the evolution form of the Cauchy-type HJ equation 
\begin{align}
\lowervalue_t(x,t) &+ \hamfunc(t; x, \nabla_x \lowervalue) = 0 \text{   in   } \Omega \times (0, T] 
\label{eq:ivp}\\
\lowervalue(x,t) &= \bm{g}, \text{ on } \partial \Omega \times \{t=T\},  \lowervalue(x, 0) = \lowervalue_0(x) \text{ in } \Omega \nonumber
\end{align}
or the convection equation 
\begin{align}
	\lowervalue_t &+ \sum_{i=0}^{N} f_i(v)_{x_i} =0, \,\, \text{ for } t > 0, \state \in \ren, \nonumber \\
	\lowervalue(\state,& 0) = \lowervalue_0(\state), \,\, \state \in \ren
\end{align}
where $\Omega$ is an open set in $\ren$; $\state$ is the state; $\lowervalue_t$ denotes the partial derivative(s) of the solution $\lowervalue$ with respect to time $t$; the Hamiltonian 
$\bm{H}: (0, T] \times \ren \times \ren \rightarrow \reline$ and $f$ are continuous; $\bm{g}, \text{   and   }  \lowervalue_0$ are bounded and uniformly continuous (BUC) functions in $\ren$; and $\nabla_x \lowervalue$ is the spatial gradient of $\lowervalue$. It is assumed that $\bm{g}$ and $\lowervalue_0$ are given. Solving problems described by \eqref{eq:ivp} under appropriate boundary and/or initial conditions using the method of characteristics is limiting as a result of crossing characteristics~\cite{Crandall1983viscosity}. In the same vein, global analysis is virtually impossible owing to the lack of existence and uniqueness of  solutions $\lowervalue \in C^1(\Omega) \times (0, T]$ even if $\hamfunc$ and $\bm{g}$ are smooth~\cite{Crandall1983viscosity}. The method of ``vanishing viscosity", based on the idea of traversing the limit as $\delta \rightarrow 0$ in the hyperbolic equation \eqref{eq:ivp} (where a parameter $\delta > 0$ ``endows" the problem in a viscosity sense as in gas dynamics~\cite{HoffDiffEq}), allows generalized (discontinuous) solutions~\cite{Evans1984} whereupon if $\lowervalue \in W_{loc}^{1,\infty}(\Omega) \times (0, T]$ and $\hamfunc \in W_{loc}^{1,\infty}(\Omega)$, one can lay claim to strong notions of  general existence, stability, and uniqueness to BUC solutions $\lowervalue^\delta$ of the  (approximate) viscous Cauchy-type HJ equation 
\begin{align}
\lowervalue_t^\delta &+ \hamfunc(t; \state, \nabla_x \lowervalue^\delta) - \delta \Delta \lowervalue^\delta = 0 \text{   in   } \Omega \times (0, T] 
\label{eq:ivp-viscous} \\
\lowervalue^\delta(x,&t) = \bm{g}, \text{ on } \partial \Omega \times \{t=T\}, \lowervalue^\delta(\state, 0) = \lowervalue_0(\state) \text{ in } \Omega  \nonumber
\end{align}
in the class $\text{BUC}(\Omega \times [0, T]) \cap C^{2,1}(\Omega \times (0, T])$ \ie continuous second-order spatial and first order time derivatives for all time $T<\infty$. 
Crandall and Lions~\cite{CrandallTwoApprox} showed that $|\lowervalue^\delta(\state,t) - \lowervalue(\state,t)| \le k\sqrt{\delta}$ for $\state \in \Omega$ and $t>0$. For most of this article, we are concerned with \textit{generalized} viscosity solutions of the manner described by \eqref{eq:ivp-viscous}.

Mitchell~\cite{Mitchell2005} connected techniques used in level set methods to reachability analysis in optimal control (that employ the viscosity solution of the HJ PDE) --- essentially showing that the zero-level set of the differential zero-sum two-person game in an  Hamilton-Jacobi-Isaacs (HJI) setting~\cite{Isaacs1965,Evans1984} constitutes the safe set of a reachability problem.  Reachability concerns evaluating the \textit{decidability} of a dynamical system's trajectories' evolution throughout a state space. Decidable reachable systems are those where one can compute all states that can be reached from an initial condition in \textit{a finite number of steps}. For $\inf$-$\sup$ or $\sup$-$\inf$ optimal control problems~\cite{LygerosReachability}, the Hamiltonian is related to the \textit{backward} reachable set of a dynamical system~\cite{Mitchell2005}.   The essential fabric of this  work revolves around reachability problems defined on co-dimension-one surfaces. This reachability theory deals with analyzing the adaptability of complex systems under uncertain dynamics to ascertain the satisfaction of all constraints. Reachability theory thus lends vast applications to various disciplines including robotics, game theory, control theory, aerospace engineering, hydrography, financial markets, biology, and economics. Problems posed in reachability contexts can be analyzed using various forms of HJ equations~\cite{Mitchell2005, ReachGaussAkametalu}. 

The well-known \texttt{LevelSet Toolbox} written by~\citet{MitchellLSToolbox} is the consolidated \texttt{MATLAB}\textregistered\ package that contains the gridding methods, boundary conditions, time and spatial derivatives, integrators and helper functions. While Mitchell motivated the execution of the toolkit in \texttt{MATLAB}\textregistered\ for the expressiveness that the  language provides, modern data manipulation and scripting libraries often render the original package non-interoperable to other libraries and packages, particularly python and its associated scientific computing libraries such as \texttt{Numpy}, \texttt{Scipy}, \texttt{PyTorch} and their variants --- libraries that are becoming prevalent in modern data analysis tooling. 

In this regard, we revisit the major algorithms necessary for implicit surface representation of HJ PDEs, write the spatial, temporal, and monotone difference schemes in Python, accelerated onto GPUs via CuPy~\cite{CUPY} and present representative numerical examples. A comparison of the performance of our library to the the original one of~\citet{MitchellLSToolbox} is then provided to show the advantages of our software. \textcolor{blue}{All the examples presented in this article conform to the IEEE 754 standard in a 64-bit base-2 format \ie FP64}. 

\section{Geometry of Implicit Surfaces and Layouts}
\label{sec:implicits}
In this section, we discuss how implicit surface functions are constructed, stored on local memory and how they are transferred to GPUs. Throughout, links to \texttt{api's}, \texttt{routines}, and \texttt{subroutines} are highlighted in \textcolor{codelblue}{\texttt{blue text}} (with a working hyperlink) and we use code snippets in Python to illustrate API calls when it's convenient. 
\begin{figure}[tb]
	\centering 
	\begin{minipage}[b]{.4\textwidth}
		\includegraphics[width=\textwidth]{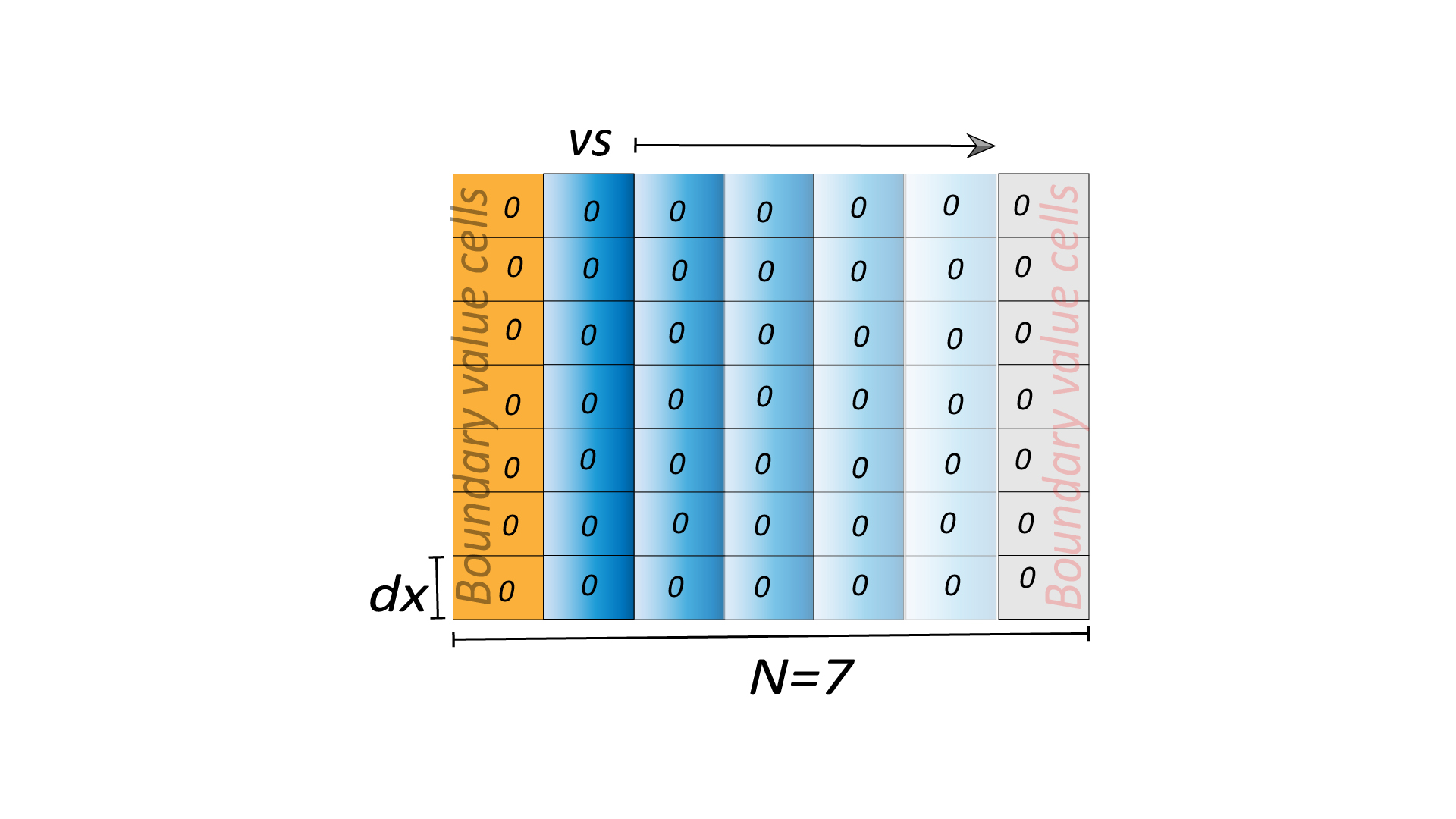}
		\subcaption{1D Grid}
	\end{minipage}
	\begin{minipage}[b]{.54\textwidth}
		\includegraphics[width=\textwidth]{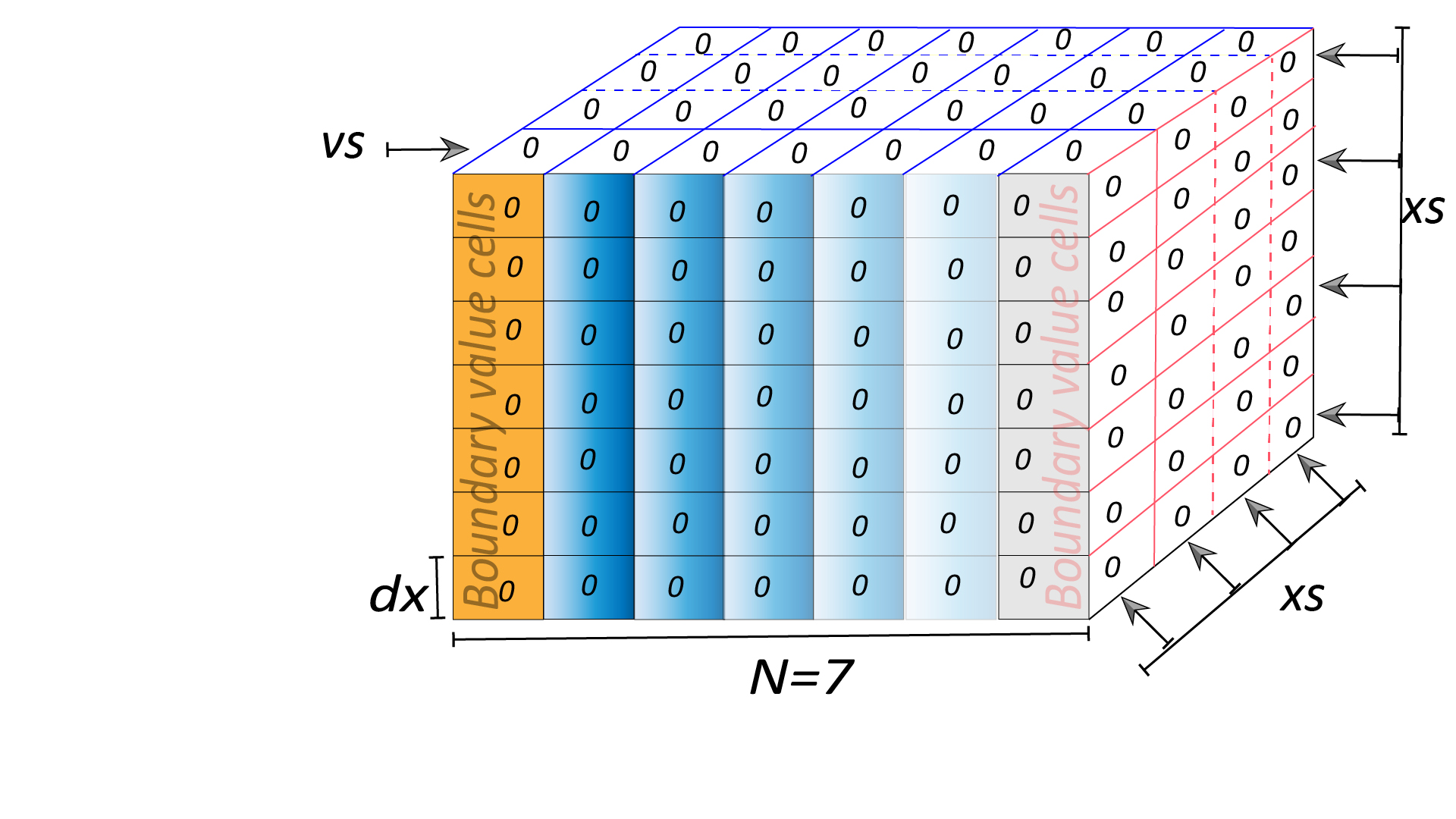}
		\subcaption{Extrapolated 1-D grid into three dimensions} 
	\end{minipage}
	\caption{\footnotesize{Creation of a grid with initial attributes $N$, $dx$, and $vs$, the total number of cells per dimension, the size of each cell, and the array of cells in each column.}}.
	\label{fig:grid_structure}
\end{figure} 

\begin{figure*}[tb!]
	\centering
	\begin{minipage}[b]{.23\textwidth}
		\includegraphics[width=\textwidth]{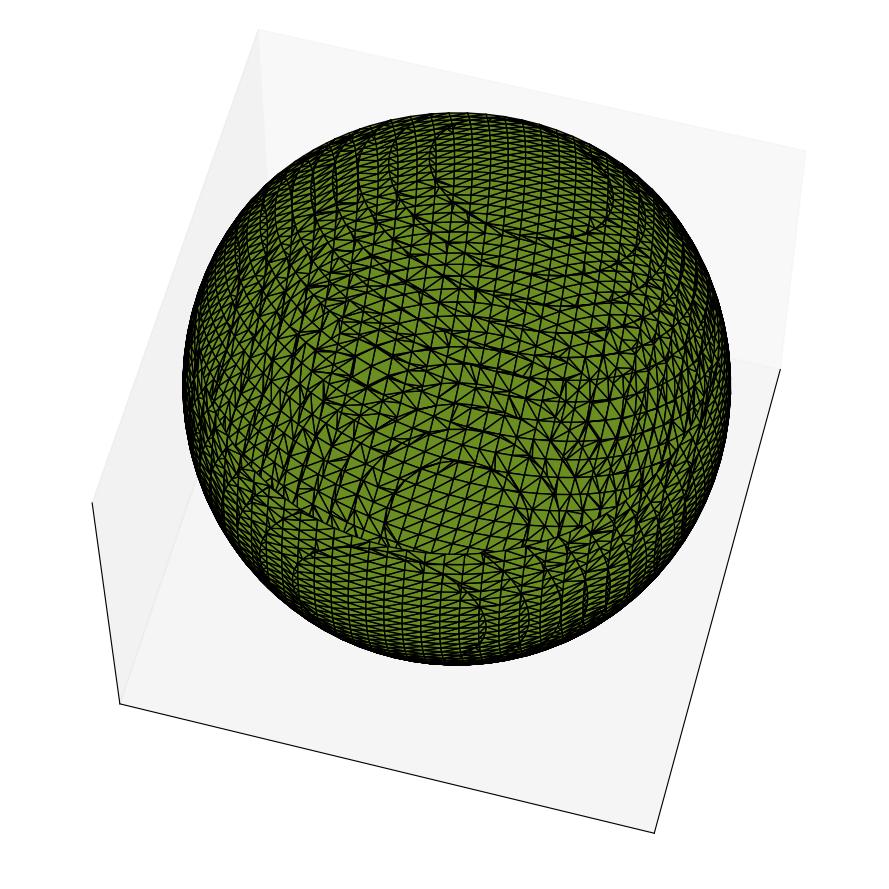}
		\subcaption{Sphere}
	\end{minipage}
	\begin{minipage}[b]{.23\textwidth}
		\includegraphics[width=\textwidth]{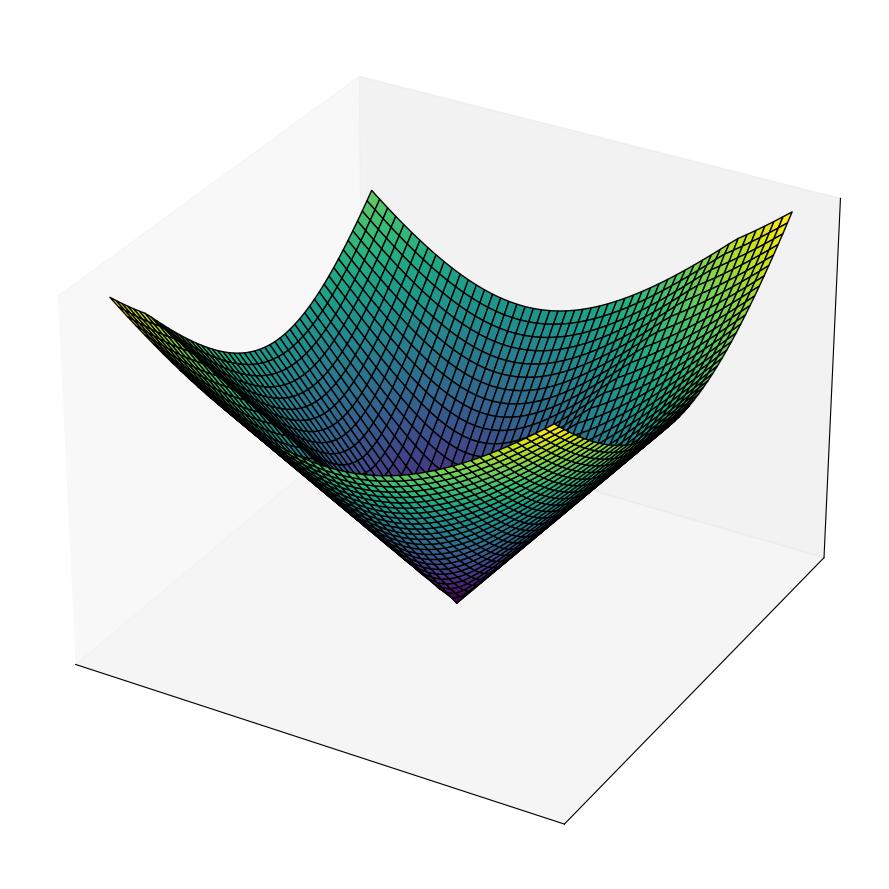} 
		\subcaption{Union of two spheres}
	\end{minipage}	
	\hfill 
	\begin{minipage}[b]{.23\textwidth}
		\includegraphics[width=\textwidth]{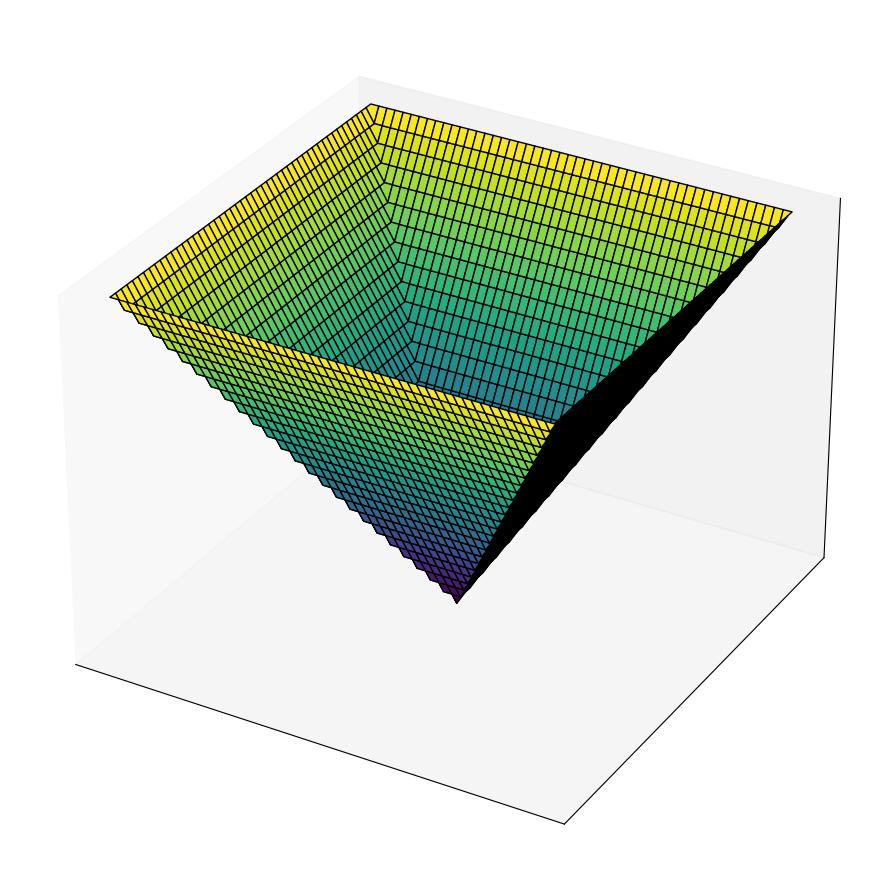}
		\subcaption{Rectangle}
	\end{minipage}
	\begin{minipage}[b]{.23\textwidth}
		\includegraphics[width=\textwidth]{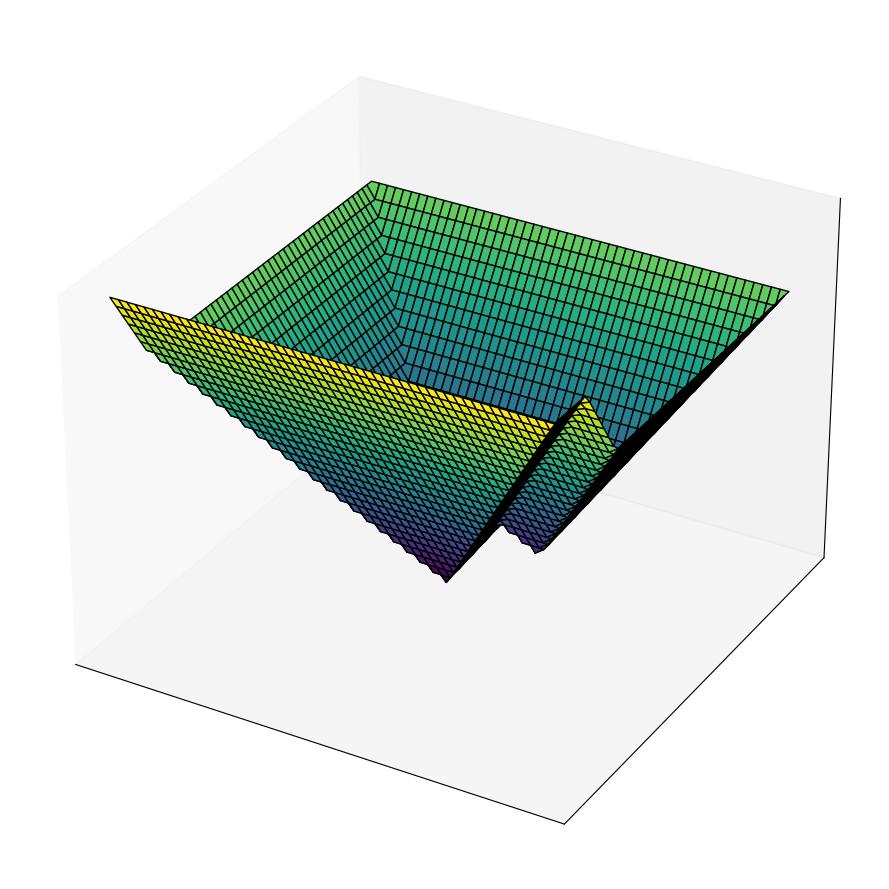} %
		\subcaption{Union of 2 rectangles}
	\end{minipage}
	%
	\caption{\footnotesize{Examples of implicitly constructed zero levelsets of interfaces of geometric primitives along with Boolean operations on 2D and 3D Cartesian grids. Zero level set of  \textbf{(a)}  a sphere on a 3D grid; \textbf{(b)} union of two 3D spheres implicitly constructed on a  2D grid; \textbf{(c)} a rectangle on a 2D grid; \textbf{(d)} the union of rectangles on a 2D grid. 
	}}
	\label{fig:implicit_3D}	
\end{figure*}

At issue are co-dimension one implicitly defined surfaces on $\mathbb{R}^n$ which represent the interface of a flow, or function e.g. $f(x)$. These interfaces are more often than not the isocontours of some function. This representation is attractive since it requires less number of points to represent a function than explicit forms. Relating to the problems of chief interest in this article, the zero isocontour (or the zero levelset)  of a reachability optimal control problem is equivalent to the safety set or backward reachable tube. Let us describe the representation of data we employ in what follows.


\subsection{Implicit surface containers}
\label{subsec:implicits::container}
Fundamental to the representation of implicit surfaces in our library are \texttt{containers} which are implemented as discrete \texttt{grid}  data structures in our library. The \texttt{grid} container is created as an associative memory that is executed as Python dictionaries of keys and corresponding values. Figure \ref{fig:grid_structure}a is an example of a  one-dimensional grid layout in memory. The corresponding keys of note in the grid data structure are the size of cells, \texttt{grid.dx} passed by the user\footnote{All cells are assumed to have a uniform size, however, a user can easily create the field \texttt{dx} as a varying-sized array of non-uniform cell sizes.}, the grid resolution, \texttt{grid.N}, which specifies the total number of grid points per dimension of $\valuefunc$ in \eqref{eq:ivp}. The attribute \texttt{grid.vs} is initialized as a one-dimensional array for all nodal points of $\valuefunc$ from \eqref{eq:ivp-viscous} in an increasingly ordered fashion from the minimum to maximum number of nodal locations. These extrema of  nodal points are specified as \texttt{min} and \texttt{max} in \texttt{grid}'s attributes. 
For problems with multidimensional $\valuefunc$, the value cells are turned into a multidimensional matrix of cells by repeating entries of \texttt{grid.vs} as row and column entries of the resulting $n-$dimensional multi-matrix of cell values of $\valuefunc$ (see \autoref{fig:grid_structure}b). The \texttt{grid.xs} attribute holds all $n-D$ matrices of coordinates of the values in \ref{eq:ivp-viscous}. It is noteworthy to remark that data is stacked in column-major format throughout the memory.
\begin{minipage}{0.95\columnwidth} 
	\centering
	\begin{lstlisting}[caption={Creating a three-dimensional grid.},label={lst:createGrid},style=pythonStyle, language=python]
		|\Hilight{pythoncolor\coloropacity}| from math import pi
		|\Hilight{pythoncolor\coloropacity}| import numpy as np
		|\Hilight{pythoncolor\coloropacity}| min = np.array((-5, -5, -pi)) // lower corner 
		|\Hilight{pythoncolor\coloropacity}|  max = np.array((5, 5, pi))   // upper corner 
		|\Hilight{pythoncolor\coloropacity}|   N = 41*ones(3, 1) // number of grid nodes
		|\Hilight{pythoncolor\coloropacity}|  pdim = 3; // periodic boundary condition, dim 3
		|\Hilight{pythoncolor\coloropacity}|   g = createGrid(min, max, N, pdim)
	\end{lstlisting}
\end{minipage}


For every dimension, at the edges of each grid layout is a placeholder array of boundary values. Boundary values are problem specific and users pass a desired boundary value call to the \texttt{grid} data structure during its creation. \autoref{fig:implicit_3D} illustrates a few implicit functions constructed on various grids. Our implementation is in the folder \href{https://github.com/robotsorcerer/levelsetpy/blob/cupy/levelsetpy/grids}{\textcolor{codelblue}{\texttt{grids}}}. 

%


\subsection{Advancing integrations via level sets}

\citet{LevelSetsBook} advocated embedding the value function on a co-dimension $R^{n-1}$ surface for an $n$-dimensional problem with signed distance functions. Any geometric primitive such as spheres, cylinders, or the like can then be appropriately constructed on the grid container of  \S \ref{subsec:implicits::container}. Specifically, for an implicit surface representation of the continuous  geometric function $\valuefunc$, we treat the coordinates of $\valuefunc$ on the state space as functional arguments instead of functional values. We do this with a fixed level set of $\lowervalue: \reline^n \rightarrow \reline$, implemented as a signed distance function from the query points of moving interfaces point sets described by implicit geometric primitives such as spheres, cylinders, or ellipsoids within the computational domain of the grid. Points inside the computational domain (or interface) possess negative values; points outside the interface possess positive values while points on the interface have zeroes assigned as their value. Then, integrating the PDE \eqref{eq:ivp-viscous} consists in moving a front throughout regions of motion in the state space by solving the levelset equation \eqref{eq:ivp-viscous} (to be described shortly).  

\begin{minipage}{0.95\columnwidth} 
	\begin{lstlisting}[caption={An \href{https://github.com/robotsorcerer/levelsetpy/blob/cupy/initialconditions/ellipsoid.py}{\texttt{ellipsoid}} as a signed distance function in our package.},label={lst:zerolevelsets},style=pythonStyle, language=python]
		|\Hilight{purple\coloropacity}| e = (g.xs[0])**2 // ellipsoid nodal points
		|\Hilight{purple\coloropacity}|  e += 4.0*(g.xs[1])**2
		|\Hilight{purple\coloropacity}|   if g.dim==3:
		|\Hilight{purple\coloropacity}\hspace{0.2cm}| data += (9.0*(grid.xs[2])**2) 
		|\Hilight{purple\coloropacity}|  e -=  radius  // radius=major axis of ellipsoid
	\end{lstlisting}
\end{minipage}

Figure \ref{fig:grid_structure} denotes a few level sets of characteristic geometric primitive implemented in our library. We describe a typical construction of an ellipsoid on a three-dimensional grid in \autoref{lst:zerolevelsets}: suppose that the zero level set of an implicit surface $\lowervalue(x,t)$ is defined as $\Gamma = \{x: \lowervalue(x)=0\}$ on a grid $G\in \ren$, where $n$ denotes the number of dimensions. Our representation of $\Gamma$ on $G$ generalizes a column-major layout. 
Level sets for several geometric primitives considered are  contained in the folder \href{https://github.com/robotsorcerer/levelsetpy/tree/cupy/levelsetpy/initialconditions}{\textcolor{codelblue}{\texttt{initialconditions}}} on our projects page.  

\section{Spatial derivatives of the value function: Upwinding}
\label{sec:upwinding}
%
Consider the levelset equation 
\begin{align}
	\lowervalue_t + \bm{F} \cdot \nabla \lowervalue = 0
	\label{eq:levelset}
\end{align} 
where $\bm{F}$ is the speed function, and the implicit function representation of $\valuefunc$ both denotes and evolves $\valuefunc$'s solution interface.  We are concerned with accurate numerical schemes for approximating the spatial derivatives to $\valuefunc$ \ie $\nabla \valuefunc$. It turns out that with essentially non-oscillatory (ENO)~\cite{ShuOsherEfficientENO} discretization schemes, $\nabla \valuefunc$ can be ``marched forward" in time such that we arrive at monotone solutions to the levelset equation \eqref{eq:levelset}. 
%
Suppose that  $\bm{F} = [f_x, f_y, f_z]$ is on a 3D Cartesian grid, expanding \eqref{eq:levelset}, we find that $\lowervalue_t + f_x \lowervalue_x + f_y \lowervalue_y + f_z \lowervalue_z = 0$ captures the implicit function's evolution on the zero levelset 
%
%
since the interface encapsulates $\lowervalue$. 
We describe the solution architecture for the  \href{https://github.com/robotsorcerer/levelsetpy/tree/cupy/spatialderivative}{\lb{\texttt{spatialderivatives}}} procedure in what follows.
%
Consider the differencing schemes on a node within a computational domain that is centered at point $i$, 
\begin{align}
D^-\lowervalue = \dfrac{\partial \lowervalue}{\partial x} \approx \frac{\lowervalue_{i+1} - \lowervalue_i}{\Delta x}, D^+\lowervalue \approx \frac{\lowervalue_{i} - \lowervalue_{i-1}}{\Delta x},
\label{eq:fwd_bck_differencing}
\end{align}
where $\Delta x$ is the (uniform) size of each cell on the grid, $\lowervalue$, and its speed $\bm{F}$ are defined over a domain $\openset$. Using the forward Euler method, the level set equation \eqref{eq:levelset} at the $n$'th integration step becomes $	(\lowervalue^{n+1}-\lowervalue^n)/\Delta t + f_x^n \lowervalue_x^n + f_y^n \lowervalue_y^n + f_z^n \lowervalue_z^n = 0.$ 
%
%
\begin{wrapfigure}{r}{0.5\textwidth}
	\begin{center}
		\includegraphics[width=0.48\textwidth]{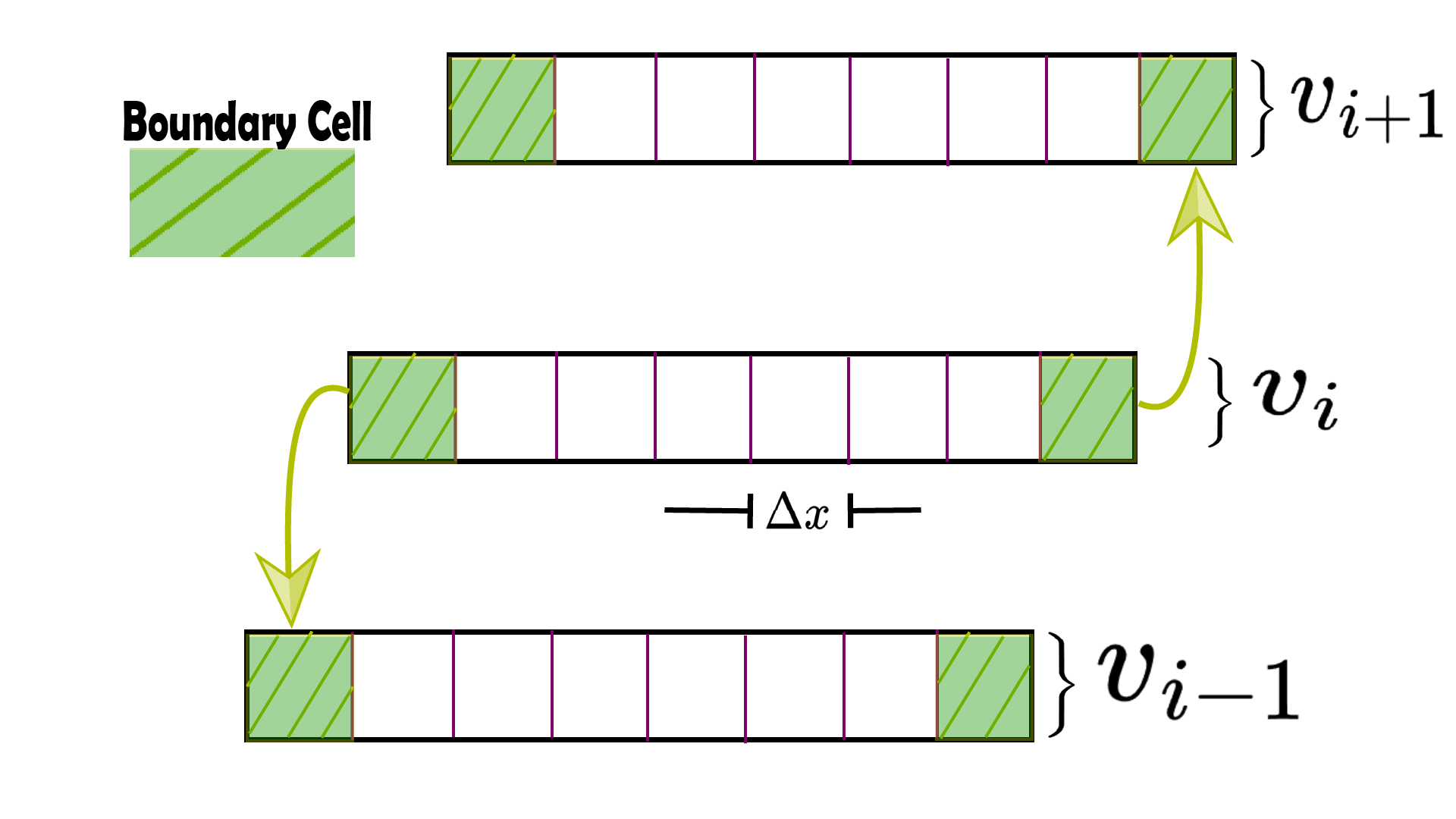}
	\end{center}
	\caption{Upwinding differencing layout.}
	\label{fig:first_upwinding}
\end{wrapfigure}
Suppose that the integration is being executed on a one-dimensional domain and around a grid point $i$, then given that $f^n$ may be spatially varying the equation in the foregoing evaluates to $(\lowervalue^{n+1}_i-\lowervalue^n_i)/{\Delta t}+ f^n_i(\lowervalue_x)^n_i  = 0$, 
%
%
where $(\lowervalue_x)_i$ denotes the spatial derivative of $\lowervalue$ w.r.t $x$ at the point $i$. We have followed the naming convention used in the MATLAB toolbox \cite{MitchellLSToolbox} and our implementation is available in the  \href{https://github.com/robotsorcerer/levelsetpy/tree/cupy/levelsetpy/spatialderivative}{\textcolor{codelblue}{\texttt{spatialderivatives}}} folder.

\subsection{First-order accurate discretization} 
\label{subsec:firstfirstENO}
Throughout, we adopt the method of characteristics~\cite[\S 3.1]{LevelSetsBook}) in choosing the direction of gradients traversal as we compute $\valuefunc_x$
: we approximate $\lowervalue_x$ with $D^-\lowervalue$ whenever $f_i > 0$ and we approximate $\lowervalue_x$ with $D^+\lowervalue$ whenever $f_i < 0$. No approximation is needed when $f_i=0$ since $f_i(\lowervalue_x)_i)$ vanishes. 
%

\begin{minipage}{0.95\columnwidth} 
	\centering
	\begin{lstlisting}[mathescape, caption={First-order accurate upwinding directional approximation to $\nabla_x \valuefunc$.},label={lst:upwindFirst},style=pythonStyle, language=python]
		|\Hilight{orgred\coloropacity}| dL, dR = upwindFirstFirst(grid, data, stencil=1) /* data $\equiv \valuefunc$ */
		|\Hilight{orgred\coloropacity}|  Host(data) $\rightarrow$ Device(data) /* transfer data to  GPU memory */
		|\Hilight{orgred\coloropacity}|   ForEach dimension $i$ of $\valuefunc$ /* stored on grid */
		|\Hilight{orgred\coloropacity}|   $\hspace{0.4cm}$ Patch data according to $\autoref{fig:first_upwinding}$  /* based on boundary value type in grid.bdry */
		|\Hilight{orgred\coloropacity}|  $\hspace{0.4cm}$  Shift data $\valuefunc_i$ to $\valuefunc_{i-1}$  and $\valuefunc_{i+1}$ according to $\autoref{fig:first_upwinding}$  
		|\Hilight{orgred\coloropacity}|  $\hspace{0.4cm}$ Obtain the left and right directional derivatives, $dL, dR$ by eq.  $\eqref{eq:fwd_bck_differencing}$
		|\Hilight{orgred\coloropacity}|  return $dL, dR$ for every other dimension of $\valuefunc$.
	\end{lstlisting}
\end{minipage}
The routine  \href{https://github.com/robotsorcerer/levelsetpy/blob/cupy/levelsetpy/spatialderivative/upwind_first_first.py}{\textcolor{codelblue}{\texttt{upwindFirstFirst}}} implements the first order derivatives \eqref{eq:fwd_bck_differencing} 
 as illustrated in \autoref{lst:upwindFirst}.
%
%

\subsection{Essentially nonoscillatory differencing}
\label{subsec:upwind_eno}
The first-order Euler derivatives computed from \S \ref{subsec:firstfirstENO} can suffer from inaccurate approximations. Stability, consistence, and convergence are necessary while numerically solving linear PDEs such as~\eqref{eq:ivp-viscous}. ~\citet{OsherShuENO} proposed using numerical flux functions of the HJ equation \eqref{eq:ivp-viscous}, calculated from a divided difference table of the nodal implicit function data. Choosing the most stable numerical flux among computed solutions constitutes the \textit{essentially non-oscillatory} (ENO) polynomial interpolant of the smooth fluxes; this preserves better accuracy as integration of the HJ equation is advanced. 

\begin{minipage}{0.95\columnwidth} 
	\centering
	\begin{lstlisting}[mathescape, caption={Second-order accurate essentially nonoscillatory upwinding approximation to $\nabla_x \valuefunc$.},label={lst:HJENO},style=pythonStyle, language=python]
		|\Hilight{light-gray\coloropacity}| dL, dR = upwindFirstENOX(grid, $\valuefunc$, $\Delta x$, stencil=2) /* $\valuefunc \equiv$ data,  $\Delta x \equiv $ cell size*/
		|\Hilight{light-gray\coloropacity}|  Host(data) $\rightarrow$ Device(data) /* transfer data to  GPU memory */
		|\Hilight{light-gray\coloropacity}|   ForEach dimension $i$ of $\valuefunc$ /* stored on grid */
		|\Hilight{light-gray\coloropacity}|   $\hspace{0.4cm}$ Patch data according to $\autoref{fig:first_upwinding}$  /* based on boundary value type in grid.bdry */
		|\Hilight{light-gray\coloropacity}|  $\hspace{0.4cm}$  Shift data $\valuefunc_i$ to $\valuefunc_{i-1}$  and $\valuefunc_{i+1}$ according to $\autoref{fig:first_upwinding}$  
		|\Hilight{light-gray\coloropacity}|  $\hspace{0.4cm}$  Set the zeroth (spatial) derivative of $\valuefunc$ as $D_i^0 \lowervalue = \lowervalue_i$ 
		|\Hilight{light-gray\coloropacity}|  $\hspace{0.4cm}$ /*Compute $\lowervalue$'s derivatives as the midway between grid nodes */
		|\Hilight{light-gray\coloropacity} | $\hspace{0.4cm}$ $D_{i+1/2}^1 \valuefunc_i = (D_{i+1}^0\valuefunc_i - D_i^0 \valuefunc_i) /{\Delta x}$ /* $1^{st}$-order divided differences of $\lowervalue$ */
		|\Hilight{light-gray\coloropacity} | $\hspace{0.4cm}$ $D_{i}^2 \lowervalue = (D_{i+1/2}^1\lowervalue - D_{i-1/2}^1 \lowervalue)/{2\Delta x}$ /* Midpoint between the grid nodes */
		|\Hilight{light-gray\coloropacity} | $\hspace{0.4cm}$ ($dL^{0,1}, dR^{0,1}) \leftarrow (D_{i+1/2}^1 \lowervalue_i^-, D_{i+1/2}^1 \lowervalue_i^+)$ /* Make two copies of left and right derivatives  */
		|\Hilight{light-gray\coloropacity} | $\hspace{0.4cm}$ ($dL^{0}, dL^{1}) += (D_{i}^2 \lowervalue_i^{-}/\Delta x, D_{i}^2 \lowervalue_{i+1}^{+}/\Delta x)$  /* Modify each $dL$ by the $2^{nd}$ order derivatives */
		|\Hilight{light-gray\coloropacity} | $\hspace{0.4cm}$ ($dR^{0}, dR^{1}) -= (D_{i}^2 \lowervalue_{i+1}^{-}/\Delta x, D_{i}^2 \lowervalue_{i+2}^{+}/\Delta x)$  /* Modify each $dR$ by the $2^{nd}$ order derivatives */
		|\Hilight{light-gray\coloropacity} | $\hspace{0.4cm}$  /* Now, we must determine the least oscillating solution */
		|\Hilight{light-gray\coloropacity} | $\hspace{0.4cm}$ $L^- = \mid D_{i-1}^2 \valuefunc \mid <  \mid D_{i+1}^2 \valuefunc \mid$, $R^- = \mid D_{i-1}^2 \valuefunc \mid >  \mid D_{i+1}^2 \valuefunc \mid$ /* Find smaller of solutions */
		|\Hilight{light-gray\coloropacity}|  $\hspace{0.4cm}$ /* Now choose approximation based on the $\min |D_2|$ value. */
		|\Hilight{light-gray\coloropacity}|  $\hspace{0.4cm}$ $dL = dL^0 * L^- + dl^1 * R^-, \quad dR = dR^0*L^-  + dR^1 * R^- $
		|\Hilight{light-gray\coloropacity}|  return $dL, dR$.
	\end{lstlisting}
\end{minipage}
The procedure for calculating the second-order accurate HJ ENO solutions as described in \cite[\S 3.3]{LevelSetsBook} is elucidated in \textcolor{blue}{\autoref{lst:HJENO}}. For third-order accurate solutions, the procedure proceeds as in \autoref{lst:HJENO} but we compute the third divided differences table as $D_{i+1/2}^3 \lowervalue = (D_{i+1}^2\lowervalue - D_i^2 \lowervalue)/{3\Delta x}$ before Line 10; the essentially non-oscillating (ENO) polynomial and its derivative are constructed as $	\lowervalue(x) = Q_0(x) + Q_1(x) + Q_2(x) + Q_3(x), \,\, 
\lowervalue_x(x_i) = Q_1^\prime(x_i) +  Q_2^\prime(x_i) + Q_3^\prime(x_i)$, with the coefficients $Q_i(x)$ and $Q_i^\prime(x)$ chosen as stipulated in~\citet[\S 3.3.]{LevelSetsBook}.
As higher order approximations may be susceptible to large gradient oscillations if the interpolant is near this neighborhood, we choose a constant $c$ such that 
$c^\star = D^3_{k^\star+1/2} \quad \text{   if   } |D^3_{k^\star+1/2} \lowervalue| \le  |D^3_{k^\star+3/2} \lowervalue|, \quad \text{ else } D^3_{k^\star+3/2} \lowervalue \quad \text{   if   } |D^3_{k^\star+1/2} \lowervalue| > |D^3_{k^\star+3/2} \lowervalue|$, where $k=i-1$ and $k=i$ for $\nabla_x \lowervalue^-$ and $\nabla_x \lowervalue^+$ respectively. 
We refer readers to the implementation on our online website: \href{https://github.com/robotsorcerer/levelsetpy/blob/cupy/levelsetpy/spatialderivative/upwind_first_eno3.py}{\textcolor{codelblue}{\texttt{upwindFirstENO3}}} within the  \href{https://github.com/robotsorcerer/levelsetpy/tree/cupy/levelsetpy/spatialderivative}{\textcolor{codelblue}{\texttt{spatialderivatives}}} folder.
\subsection{Weighted Essentially Nonoscillatory HJ Solutions}
Towards improving the accuracy of first-order accurate monotone schemes for the viscosity solution to \eqref{eq:ivp},~\citet{WeightedENOPengJiang} proposed weighted ENO (WENO) schemes which weights substencils of the base ENO stencil according to the solution's relative smoothness  on the chosen substencils. 
%

Our  implementation of the WENO scheme leverages the computed ENO schemes of \S \ref{subsec:upwind_eno} in addition to computing the smoothness coefficients and weights according to the recommendation in~\citet{ShuOsherEfficientENO}.  Our routine is available in \href{https://github.com/robotsorcerer/levelsetpy/blob/cupy/levelsetpy/spatialderivative/upwind_first_weno5a.py}{\textcolor{codelblue}{\texttt{upwindFirstWENO5a}}} and called as \href{https://github.com/robotsorcerer/levelsetpy/blob/cupy/levelsetpy/spatialderivative/upwind_first_weno5.py}{\textcolor{codelblue}{\texttt{upwindFirstWENO5}}}.  

\section{The Hamiltonian: A Lax-Friedrichs Approximation Scheme}
\label{ssec:lax-friedrichs}
Hyperbolic Hamilton-Jacobi partial differential equations such as
 \begin{align}
	\valuefunc_t + H(t; x, \nabla_x \valuefunc)  = 0
	\label{eq:HJEquation}
\end{align} 
where $H(\nabla \valuefunc)$ is the Hamiltonian are a form of conservative laws which occur in many natural processes. In 3D, they can be written as $\valuefunc_t + H(\nabla_x \valuefunc, \nabla_y \valuefunc, \nabla_z \valuefunc)  = 0$. Resolving this HJ PDE is impossible analytically; however, one can  numerically discretize it 
so that its numerical approximation becomes $\hat{H}(\nabla \valuefunc^-, \nabla \valuefunc^+)$ or $\hat{H}(\nabla_x \valuefunc^-, \nabla_x \valuefunc^+, \nabla_y \valuefunc^-, \nabla_y \valuefunc^+, \nabla_z \valuefunc^-, \nabla_z \valuefunc^+)$ in higher dimensions. Then leveraging the classical equivalence between the solutions to scalar conservation laws and HJ equations in one spatial dimension, a numerically \textit{consistent} solution to $H$ can be found such that $H(\nabla \valuefunc) \equiv \hat{H}(\nabla \valuefunc)$.

The spatial derivatives in $\hat{H}$ can be approximated with one of the upwinding schemes of \S \ref{sec:upwinding}. To discretize ${H}$, the Lax-Friedrichs scheme~\cite{CrandallLaxFriedrichs} comes into play. Numerical discretization of the HJ equation proceeds as 
\begin{align}
	(\valuefunc^{n+1} - \valuefunc^n)/\Delta t + \hat{H}^n(\nabla_x \valuefunc^-, \nabla_x \valuefunc^+, \nabla_y \valuefunc^-, \nabla_y \valuefunc^+, \nabla_z \valuefunc^-, \nabla_z \valuefunc^+) = 0
\end{align}
and \textit{consistency} in the numerical Hamiltonian's solution with the actual system Hamiltonian is enforced via a Courant-Friedreichs-Lewy (CFL) condition which states that the speed of a physical wave be no faster than a numerical wave. Put differently, we require $|\valuefunc| > \Delta x / \Delta t$ (for a single-dimensional system). In this sentiment, the Lax-Friedrichs approximation scheme for $H$ in 2D is of the form 
\begin{align}
	\hat{H} = H\left(\dfrac{\nabla_x \valuefunc^+ + \nabla_x \valuefunc^-}{2}, \dfrac{\nabla_y \valuefunc^+ + \nabla_y \valuefunc^-}{2}\right) - \alpha_x \left(\dfrac{\nabla_x \valuefunc^+ - \nabla_x \valuefunc^-}{2}\right)  - \alpha_y \left(\dfrac{\nabla_y \valuefunc^+ - \nabla_y \valuefunc^-}{2}\right) 
\end{align}
where the quantity of numerical viscosity is managed by the dissipation coefficients $\alpha_x, \alpha_y$ defined as $\alpha_x = \max|H_x(\valuefunc_x, \valuefunc_y)|, \,\, \alpha_y = \max |H_y(\valuefunc_x, \valuefunc_y)|$. To preserve stability of the integration scheme, we follow the CFL stability consition of \citet[\S 5.3]{LevelSetsBook} i.e., $\Delta t  \max \{| H_x | /\Delta x +  | H_y | /\Delta y + | H_z | /\Delta z \} < 1$.

\begin{minipage}{0.95\columnwidth} 
	\centering
	\begin{lstlisting}[mathescape, caption={Computational scheme for the Lax-Friedrichs Approximation to $Grad H(x, p)$.}, style=pythonStyle, language=python, label={lst:laxFriedrichs}]
		|\Hilight{pythoncolor\coloropacity}|  termLaxFriedrichs($t_{init}, v$, sysData) /* sysData struct:  grid info, $p, \alpha, \hat{H}, \nabla \hat{H}$ routines */
		|\Hilight{pythoncolor\coloropacity}| dL, dR, dC = sysData.CoStateCalc(sysData.grid, $\valuefunc$) /* Obtain co-state i.e. left, 
		|\Hilight{pythoncolor\coloropacity}|  /* right, and centered derivatives using one of the upwinding schemes */; 
		|\Hilight{pythoncolor\coloropacity}|   $\hat{H}$ =  sysData.hamFunc($t_{init}, \valuefunc, dC$, sysData) /* compute the numerical Hamiltonian */
		|\Hilight{pythoncolor\coloropacity}| $\alpha, \Delta t \, \max \{| H |/\Delta x \}$ = sysData.dissFunc($t_{init}, \valuefunc, dL, dR$, sysData) /* Lax-Friedrichs
		|\Hilight{pythoncolor\coloropacity}| /* stabilization coefficients and CFL condition */
		|\Hilight{pythoncolor\coloropacity}|   $\nabla \hat{H} = \hat{H}  - \alpha$ /* update the implicitly stored data based on the change. */
	\end{lstlisting}
\end{minipage}
Our Lax-Friedrichs  implementation scheme is described in the pseudo-code of \autoref{lst:laxFriedrichs}. It takes as input the initial time step for the current integration horizon, the value function data embedded within the grid structure, and a \texttt{sysData} data structure that consists of the routines for computing the user-supplied problem's co-state, $p$, dissipation coefficients (\ie $\alpha_x$, $\alpha_y$), and a function for evaluating the numerical Hamiltonian. The spatial derivatives of $\valuefunc$ are computed by choosing a desired routine from one of the upwinding schemes described in \S \ref{sec:upwinding} which are then passed to the internal numerical Hamiltonian calculator as seen on Line 4. For details, we refer readers to the online implementation on this \href{https://github.com/robotsorcerer/levelsetpy/blob/cupy/levelsetpy/explicitintegration/term/term_lax_friedrich.py}{\textcolor{codelblue}{\texttt{link}}}. 
\section{The HJ Equation: Total Variation Diminishing Runge-Kutta}
\label{sec:temporal}
So far we have considered numerical methods for solving the spatial derivatives $\nabla_x \valuefunc$ and the numerical Hamiltonian $\hat{H}(t, x, \nabla_x \valuefunc)$ in the hyperbolic HJ PDE \eqref{eq:HJEquation}. We now shift our attention to spatial discretization methods for resolving the hyperbolic PDE \eqref{eq:HJEquation} altogether. Ours is an implementation of the non-oscillatory total variation diminishing (TVD) spatial discretization CFL-constrained Runge-Kutta (RK) scheme via method of lines~\cite{ShuOsherEfficientENO, ShuOsherEfficientENOII}. Under TVD-RK schemes in general, it is required that the total variation of the numerical solution to $\valuefunc$ \ie $TV(\valuefunc) = \sum_j |\valuefunc_{j+1} - \valuefunc_j|$ does not increase $TV(\valuefunc^{n+1}) \le TV(\valuefunc^n)$ for all $n$ and $\Delta t$ such that $0 \le n \Delta t \le T$ for first-order in time Euler forward integrators. To maintain the TVD property, it is typical to impose a CFL time step constraint such as $\Delta t \le c \Delta x$, where $c$ is a CFL coefficient for the higher-order time discretization. Following~\citet{ShuOsherEfficientENO} observations about the limitations of higher-order schemes with wider stencils that cause numerical degradation, we limit our implementations to the recommended third-order schemes which are known to work well. For the technical details behind our implementation, we refer readers to~\citet{ShuOsherEfficientENO}. 

\begin{minipage}{0.95\columnwidth} 
	\centering
	\begin{lstlisting}[mathescape, caption={A $3^{rd}$-order TVD-RK Scheme for integrating an hyperbolic HJ PDE by MoL.},label={lst:odeCFL},style=pythonStyle, language=python]
		|\Hilight{codegreen\coloropacity}| t, $\valuefunc$ = odeCFL3(func, tspan, $\valuefunc_0$) /* func:= Lax-Friedrichs routine */
		|\Hilight{codegreen\coloropacity}|  Host($\valuefunc_0$) $\rightarrow$ Device($\valuefunc_0$) /* transfer data to  GPU memory */
		|\Hilight{codegreen\coloropacity}|   $\valuefunc \leftarrow \valuefunc_0$  /* Copy over initial value function */
		|\Hilight{codegreen\coloropacity}| cfl_factor = 0.32; stepBound = 0
		|\Hilight{codegreen\coloropacity}|  while $t_f - t \ge \epsilon t_f$ /* $t_f =$ final time in tspan; $\epsilon > 0$  */ 
		|\Hilight{codegreen\coloropacity}|  $\hspace{0.4cm}$ ForEach $\valuefunc_{idx}$ in $\valuefunc$  
		|\Hilight{codegreen\coloropacity}|  $\hspace{0.6cm}$  $\dot{\valuefunc}_{idx}, \, t_{idx}$ = func(t, $\valuefunc$) /* Compute the numerical approximation of $H(x,p)$ */
		|\Hilight{codegreen\coloropacity}|  $\hspace{0.6cm}$ $\Delta t = \min(t_f-t$, maxStep, cfl_factor*stepBound) /* CFL timestep bound, up to $t_f$.*/
		|\Hilight{codegreen\coloropacity}| $\hspace{0.6cm}$ /* First substep */
		|\Hilight{codegreen\coloropacity}| $\hspace{0.6cm}$  $t_1 = t + \Delta t $,  $\valuefunc_1 = \valuefunc + \Delta t * \dot{\valuefunc}$ 
		|\Hilight{codegreen\coloropacity}| $\hspace{0.6cm}$  /* Second substep */
		|\Hilight{codegreen\coloropacity}| $\hspace{0.6cm}$  $\dot{\valuefunc}_{idx}, \, t_{idx}$ = func($t_1, \valuefunc_1$) /*  Euler forward stepping $t_{n+1} \rightarrow  t_{n+2}$*/
		|\Hilight{codegreen\coloropacity}| $\hspace{0.6cm}$  $t_2 = t_1 + \Delta t; \,\, \valuefunc_2 = \valuefunc_1 + \Delta t * \dot{\valuefunc}$  /* Update time step and values */
		|\Hilight{codegreen\coloropacity}| $\hspace{0.6cm}$  $t_{1/2} =  (3t + t_2)/4, \quad \valuefunc_{1/2} = (3 \valuefunc + \valuefunc_2)/4$ /* approximation at $t_{n+1/2}$ */
		|\Hilight{codegreen\coloropacity}| $\hspace{0.6cm}$  /* Third substep */
		|\Hilight{codegreen\coloropacity}| $\hspace{0.6cm}$  $\dot{\valuefunc}_{idx}, \, t_{idx}$ = func($t_{1/2}, \valuefunc_1{1/2}$) /*  Euler forward stepping   $t_{n+1} \rightarrow  t_{n+2}$*/
		|\Hilight{codegreen\coloropacity}| $\hspace{0.6cm}$  $t_{3/2} = t_{1/2} + \Delta t; \,\, \valuefunc_{3/2} = \valuefunc_{1/2} + \Delta t * \dot{\valuefunc}$  /* Update time step and values */
		|\Hilight{codegreen\coloropacity}| $\hspace{0.6cm}$  $t  =  (t + 2t_{3/2})/3, \quad \valuefunc = (\valuefunc + 2 \valuefunc_{3/2})/3$ /* approximation at $t_{n+1}$ */
		|\Hilight{codegreen\coloropacity}| return $t, \valuefunc$
	\end{lstlisting}
\end{minipage}
\autoref{lst:odeCFL} describes a representative implementation of the method of lines integration scheme. We refer readers to the \href{https://github.com/robotsorcerer/levelsetpy/blob/cupy/explicitintegration/integration/ode_cfl_3.py}{\textcolor{codelblue}{\texttt{integration}}} folder for all of our implementations.
\begin{wrapfigure}{r}{0.5\textwidth}
	\centering
	\includegraphics[height=.4\textwidth,width=.35\columnwidth]{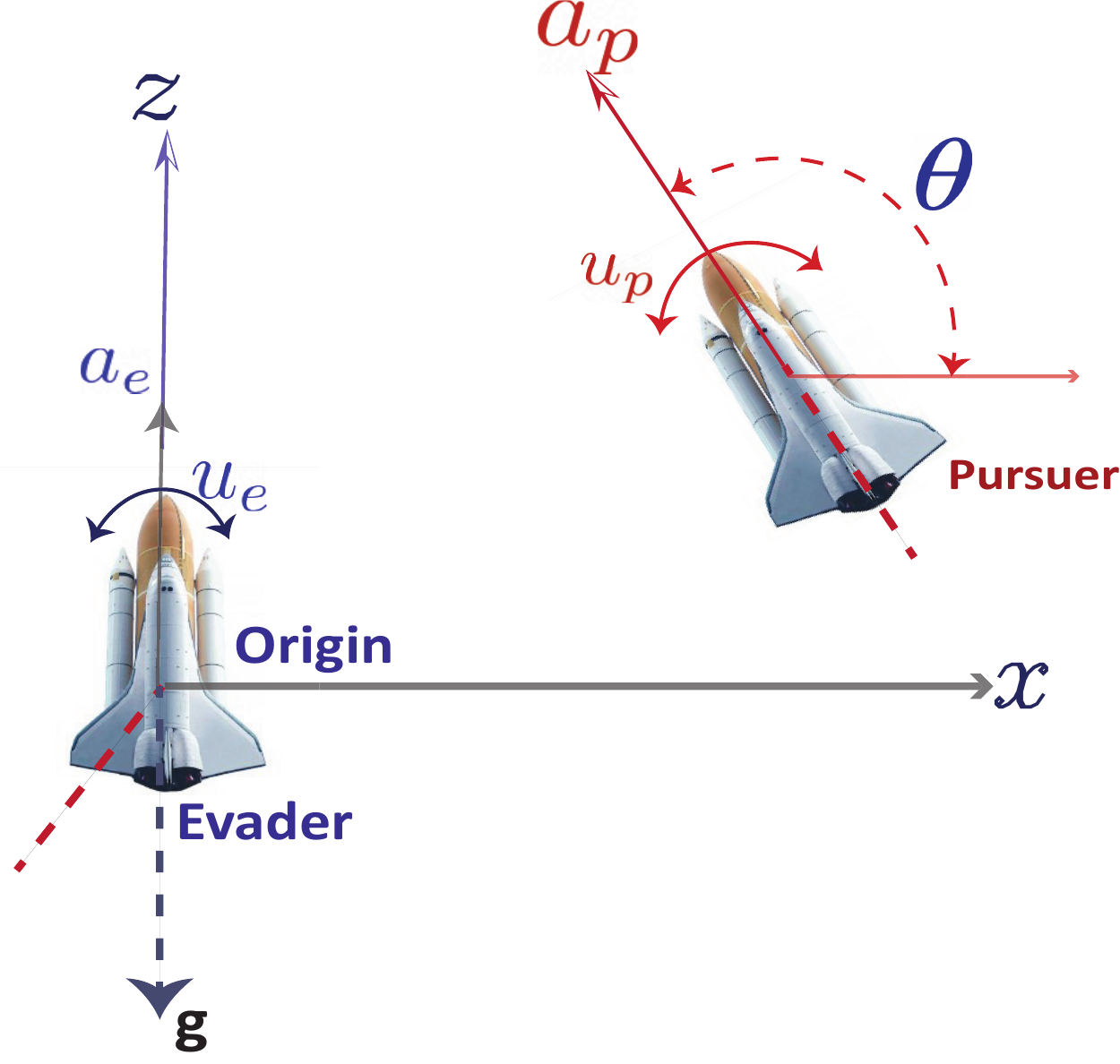}
	\caption{Two rockets on a Cartesian $\bm{xz}$-plane with relative thrust inclination $\theta:=u_p- u_e$.}
	\label{fig:rocket_relative}
\end{wrapfigure}
\section{Numerical Experiments}
\label{sec:results}
In this section, we will present problems motivated by real-world scenarios and amend them to HJ \pde forms where their numerical solutions can be resolved with our \texttt{levelsetpy} toolbox. \textcolor{red}{W \textcolor{blue}{All the examples presented in conform to the FP64 double floating precision format}}.
We consider \textit{differential games} as a \textit{collection/family of games}, $\Upsilon = \{\Gamma_1, \cdots, \Gamma_g\}$ where each game may be characterized as a pursuit-evasion \textit{game}, $\Gamma$.  Such a game terminates when \textit{capture} occurs, that is the distance between players falls below a predetermined threshold. Each player in a game shall constitute either a pursuer ($\pursuer$) or an evader ($\evader$). 
The essence of our  examples is to  geometrically (approximately) ascertain the separation between the \textit{barrier hypersurface}, where starting points exist for which escape occurs, capture occurs, and for which the outcome is neutral.
The task is to assay the \textit{game of kind} for the envelope of the capturable states. This introduces the \textit{barrier} hypersurface which separates, in the initial conditions space, the hypersurface of capture from those of escape. In this \textit{game of kind} postulation, all optimal strategies are not unique, but rather are a \textit{legion}. Ergo, we are concerned with the set of initial positions on the vectogram where the capture zone (CZ) exists \ie where game termination occurs; and the nature of escape zones (EZ) \ie zones where termination or escape does not occur -- after playing the differential game. 

\subsection{The Rockets Launch Problem}
\label{subsec:rockets_launch}
We consider the rocket launch problem of Dreyfus~\cite{Dreyfus1966} and amend it to a differential game between two identical rockets, $\pursuer$ and $\evader$, on an $(x,z)$ cross-section of a Cartesian plane. 
We want to compute  the backward reachable tube (BRT)~\cite{Mitchell2005} of the \textit{approximate} terminal surface's boundary for a predefined target set over a time horizon (\ie the target tube).  
The BRT entails the state-space regions for which min-max operations over either \textit{strategy} of $\pursuer$ or $\evader$ is below zero, and where the \textit{HJI PDE's value functional} is exactly zero.  

For a two-player differential game, let $\pursuer$ and $\evader$ share identical dynamics in a general sense so that we can freely choose the coordinates of $\pursuer$; however, $\evader$'s origin is a distance $\payoff$ away from $(x,z)$ at plane's origin (see Fig. \ref{fig:rocket_relative}) so that the $\pursuer\evader$ vector's inclination measured counterclockwise from the $\state$ axis is $\theta$.

Let the states of $\pursuer$ and $\evader$  be denoted by $(\state_p, \state_e)$. Furthermore, let the $\pursuer$ and $\evader$ rockets be driven by their thrusts, denoted by $(u_p, u_e)$ respectively  (see Figure \ref{fig:rocket_relative}). Fix the rockets' range so that what is left of the motion of either $\pursuer$ or $\evader$'s  is restricted to orientation on the $(x,{z})$ plane as illustrated in \autoref{fig:rocket_relative}. It follows that the relevant \textit{kinematic equations} (KE) (derived off \cite{Dreyfus1966}'s single rocket dynamics) is
	\begin{align} 
		\dot{x}_{2e} &= x_{4e}; \,\,  \dot{x}_{2p} = x_{4p}; \,\,
		\dot{x}_{4e} = a \sin u_e - g; \,\, \dot{x}_{4p} = a \sin u_p - g
		\label{eq:dreyfus_mitter_relevant_eq}
	\end{align}
\noindent where $a$ and $g$ are respectively the acceleration and gravitational accelerations (in feet per square second)\footnote{We set $a=1 ft/sec^2$ and $g=32 ft/sec^2$ in our simulation.}.  
$\pursuer$'s motion relative to $\evader$'s  along  the $(\state,\bm{z})$ plane includes the relative orientation shown in \autoref{fig:rocket_relative} as $\theta=u_p- u_e$, the control input. Following the conventions in \autoref{fig:rocket_relative}, the game's relative equations of motion in \textit{reduced space}~\cite[\S 2.2]{Isaacs1965} \ie
is $\bm{x} = (x, {z}, {\theta})$ where ${\theta }\in \left[-\frac{\pi}{2}, \frac{\pi}{2}\right)$ and $(x,z) \in \bb{R}^2$ are $\dot{\state} = \{\dot{x}, \dot{z}, \dot{\theta}\}$, and
%
\begin{align}
		\dot{x} = a_p \cos \theta + u_e x, \,\,\dot{z} =a_p \sin \theta + a_e + u_e x - g, \,\, \dot{\theta} = u_p -u_e.
	\label{eq:rocket_me}
\end{align}

The capture radius of the origin-centered circle $\payoff$ (we set $\payoff=1.5$ ft) is $\|\pursuer \evader\|_2 $ so that $\payoff^2 =  x^2 + z^2.$
%
All capture points are specified by  the variational HJ \pde~\cite{Mitchell2005}: 
\begin{align}
	\dfrac{\partial \payoff}{\partial t} (\bm{x},t) + \min \left[0, \hamfunc(\state, \frac{\partial \payoff(\state, t)}{\partial \state})\right] \le 0,
\end{align}
with Hamiltonian given by
\begin{align}
	\hamfunc(\state, p) &= -\max_{u_e \in [\underline{u}_e, \bar{u}_e]} \min_{u_p \in [\underline{u}_p, \bar{u}_p].
	} \begin{bmatrix}
		p_1 & p_2 & p_3
	\end{bmatrix} 
	\begin{bmatrix}
		a_p \cos \theta + u_e x \\
		a_p \sin \theta + a_e + u_p x - g 	u_p -u_e
	\end{bmatrix}.
	\label{eq:ham_def}
\end{align}
Here,  $p$ are the co-states, and $[\underline{u}_e, \bar{u}_e]$ denotes extremals that the evader must choose as input in response to the extremal controls that the pursuer plays \ie $[\underline{u}_p, \bar{u}_p]$. 
%
Rather than resort to analytical \textit{geometric reasoning}, we may analyze possibilities of behavior by either agent via a principled numerical simulation. This is the essence of this work. 
From \eqref{eq:ham_def}, set $\underline{u}_e = \underline{u}_p = \underline{u} \triangleq -1$ and $\bar{u}_p = \bar{u}_e = \bar{u} \triangleq +1$ so that $\hamfunc(\state, p)$ is
\begin{align}
	\hamfunc(\state, p)=& -\max_{u_e \in [\underline{u}_e, \bar{u}_e]} \min_{u_p \in [\underline{u}_p, \bar{u}_p]
	} 
	\begin{bmatrix}
		p_1(a_p \cos \theta + u_e x) +  p_2 (a_p \sin \theta + a_e +  u_p x - g) + p_3 (u_p -u_e)
	\end{bmatrix}, \nonumber \\
	&\triangleq -a p_1 \cos \theta - p_2 (g - a - a\sin \theta) -\bar{u} |p_1 x +  p_3 | + \underline{u} |p_2 x + p_3 |,
	\label{eq:rocket_hamfunc}
\end{align}
where the last line follows from setting $a_e = a_p \triangleq a$.

For the target set guarded by $\evader$, we choose an implicitly constructed cylindrical mesh on a three-dimensional grid. The grid's nodes are uniformly spaced apart at a resolution of $100$ points per dimension over the interval $[-64, 64]$. In numerically solving for the Hamiltonian  \eqref{eq:rocket_hamfunc}, a TVD-RK discretization scheme~\cite{OsherShuENO} based on fluxes is used in choosing smooth nonoscillatory results as described in \S \ref{sec:temporal}. Denote by $(x, y, z)$ a generic point in $\bb{R}^3$ so that given mesh sizes $\Delta x, \, \Delta y, \, \Delta z, \, \Delta t \, > 0$, letters $u,v,w$ represent functions on the $x,y,z$ lattice: $\Delta=\{(x_i,y_j, z_k): i, j, k \in \bb{Z}\}$. 
\begin{figure*}[tb!]
	\centering
	\begin{minipage}[tb]{.31\textwidth}
		\includegraphics[width=\textwidth]{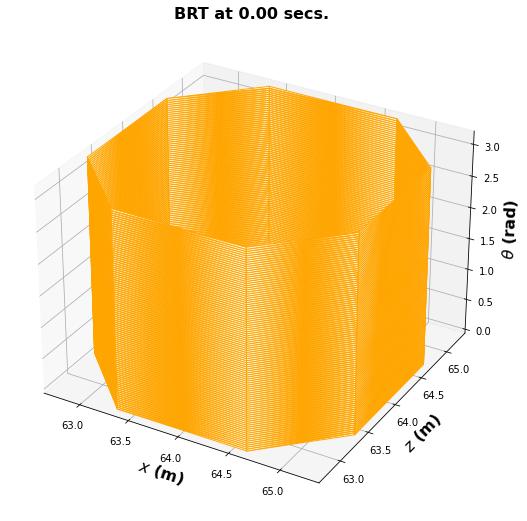}
	\end{minipage}
	\hfill
	%
	\begin{minipage}[tb]{.31\textwidth}
		\includegraphics[width=\textwidth]{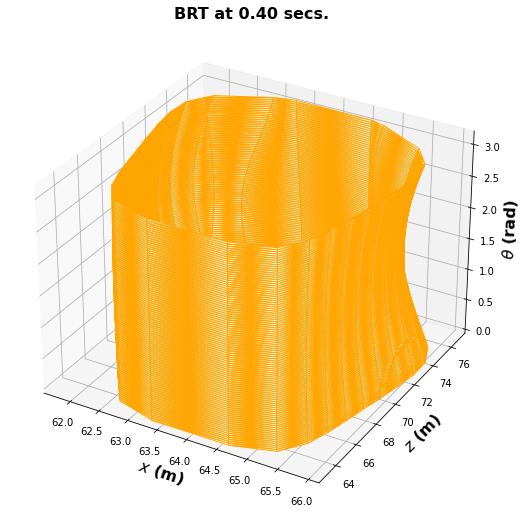}
	\end{minipage}
	\hfill
	\begin{minipage}[tb]{.31\textwidth}
		\includegraphics[width=\textwidth]{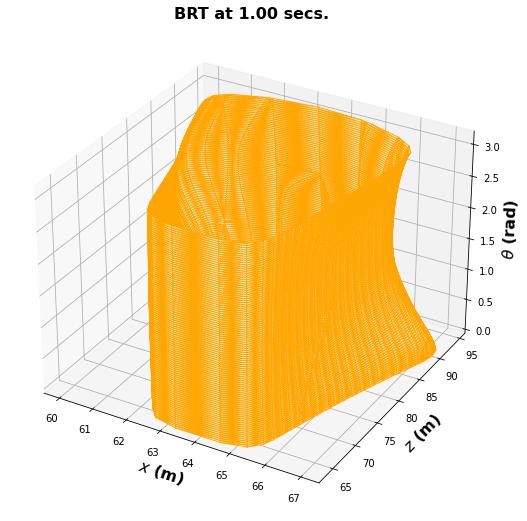}
	\end{minipage}
	\caption{\footnotesize{\textit{(Left to Right)}: Backward reachable tubes  (capture surfaces) for the rocket system (\cf \autoref{fig:rocket_relative}) optimized for the paths of slowest-quickest descent in equation \eqref{eq:ham_def} at various time steps during the differential game. 
			In all, the BRTs were computed using the method outlined in \cite{Crandall1984, LevelSetsBook, Mitchell2020}. We set $a_e = a_p = 1ft/sec^2$ and $g=32 ft/sec^2$ as in Dreyfus' original example. }}
	\label{fig:rockets_results}
\end{figure*}

\begin{minipage}{0.95\columnwidth} 
	\centering
	\begin{lstlisting}[mathescape, caption={HJ ENO2 computational scheme for the rockets.}, label={lst:rockets},style=pythonStyle, language=python]
	|\Hilight{pythoncolor\coloropacity}|finite_diff_data = {"innerFunc": termLaxFriedrichs,
	|\Hilight{pythoncolor\coloropacity}|$\hspace{0.8cm}$ "innerData": {"grid": g, "hamFunc": rocket_rel.ham,
	|\Hilight{pythoncolor\coloropacity}|$\hspace{0.8cm}$ "partialFunc": rocket_rel.dissipation,
	|\Hilight{pythoncolor\coloropacity}|$\hspace{0.8cm}$ "dissFunc": artificialDissipationGLF,
	|\Hilight{pythoncolor\coloropacity}|$\hspace{0.8cm}$ "CoStateCalc": upwindFirstENO2},
	|\Hilight{pythoncolor\coloropacity}|$\hspace{0.8cm}$ "positive": True}  // direction of approx. growth 
	\end{lstlisting}
\end{minipage}

The Hamiltonian, upwinding scheme, flux dissipation method, and the overapproximation parameter for the essentially non-oscillatory polynomial interpolatory data used in geometrically reasoning about the \textit{target tube} is set up as seen in \autoref{lst:rockets}. The data structure \texttt{finite\_diff\_data} contains all the routines needed for adding dynamics to the original implicit surface representation of $\lowervalue(\state, t)$. The  Hamiltonian is approximated with the routine \texttt{termLaxFriedrichs} described in \S \ref{ssec:lax-friedrichs}), and  passed to the \texttt{innerFunc} query field. During integration, it is calculated as derived in \eqref{eq:rocket_hamfunc} within the \texttt{hamFunc} query field. We adopt a second-order accurate essentially nonoscillatory HJ upwinding scheme is used to compute $\nabla \valuefunc$. Since we are overapproximating the value function, we specify a \texttt{True} parameter for the \texttt{positive} query field.

Safety is engendered by having the evader respond optimally to the pursuer at various times during the game. The entire safety set over the time interval of play constitutes the backward reachable tube (BRT)~\cite{Mitchell2005}; this BRT under the control strategies of $\pursuer$ or $\evader$, is a part of the phase space, $\openset \times T$. We compute the \textit{overapproximated} BRT of the game over a time span of $[-2.5, 0]$ seconds during 11 global optimization time steps using the CFL-constrained TVD-RK solver for the HJ equation..  

The initial value function (leftmost inset of \autoref{fig:rockets_results}) is represented as a (closed) dynamic implicit surface over all point sets in the state space (using a signed distance function) for a coordinate-aligned cylinder whose vertical axes runs parallel to the rockets'  orientation (see  \autoref{fig:rocket_relative}). This closed and bounded assumption of the target set is a prerequisite of backward reachable analysis (see ~\cite{Mitchell2005}). It allows us to include all limiting velocities.  We reach convergence at the eleventh global optimization timestep (rightmost inset of \autoref{fig:rockets_results}).  

Reachability~\cite{Mitchell2001, LygerosReachability} thus affords us an ability to numerically reason about the behavior of these two rockets before motion execution without closed-form geometrical analysis. This example demonstrates running a CFL-restricted TVD-RK optimization scheme for a finite number of global optimization timesteps. More complicated examples are available on our github repository and we encourage readers to use and test the library for multiple other safety analysis and/or synthesis.

\subsection{Computational Time Comparison with LevelSet Toolbox}
We compare the solution for recovering the zero level set of the system presented in the previous example against~\citet{MitchellLSToolbox}'s \texttt{LevelSet Toolbox}. Additional results are available within this article's accompanying supplementary materials, and in the \href{https://github.com/robotsorcerer/levelsetpy}{\textcolor{blue}{online package}}. In all, we compare the efficacy of running various computational problems using our library on a \texttt{CPU} -- running \texttt{Numpy} versus  \texttt{MATLAB's}\textregistered levelset toolbox~\cite{MitchellLSToolbox} -- and on a \texttt{GPU}. \textcolor{blue}{The CPU examples all run on the main thread in order for us to provide a measurable comparison to the original MATLAB implementation. We do not leverage any multiple thread computational tools such as OpenMP or Pthread. Throughout the code, there are parts that execute on the host and the parts that consume a lot of memory are transferred to the device. We do this by design so that we do not create a communication bottleneck in computation between computational operations that require a  light memory consumption footprint and large computational operations that could benefit from device computation.} For the \texttt{CPU} tests, we run the computation on an  Intel Core\texttrademark i9-10885H  16 cores-processor with a  2.4GHz clock frequency, and 62.4GB memory. On \texttt{GPUs}, we employed an \texttt{NVIDIA Quadro RTX} 4000 with $8.192$ GiB memory running on a mobile workstation in all of our \texttt{GPU} library accelerations.
\begin{table}[tb!]
	\caption{Time to Resolve HJ \pde's. \textcolor{blue}{N/A signifies the experiment was not executed in  \textsc{MATLAB} due to its memory inefficient programmatic scheme.}}
	\label{tab:time_compute}
	\begin{tabular}{|p{2.2cm}|c|p{1.2cm}|c|p{1.2cm}|c|p{1.2cm}|r|}
		\hline
		\multirow{2}{*}{\backslashbox{Expt}{Lib}}  & \multicolumn{2}{|c|}{\footnotesize{levelsetpy GPU Time (secs)}} & \multicolumn{2}{|c|}{\footnotesize{levelsetpy CPU Time (secs)}} & \multicolumn{2}{|c|}{\footnotesize{MATLAB CPU (secs)}} \\ 
		~  & \footnotesize{Global} & \footnotesize{Avg. local} & \footnotesize{Global} & \footnotesize{Avg. local} & \footnotesize{Global} & \footnotesize{Avg. local} \\ \hline 
		Rockets & $11.5153\pm 0.038$  & $1.1833$ & $107.84 \pm 0.42$ & $10.4023$  &  $138.50$ & $13.850$   \\ \hline
		Doub. Integ.   &  $14.7657 \pm 0.2643$  & $1.5441$ & $3.4535 \pm 0.34$ & $0.4317$ &  $5.23$ &$0.65375$ \\ \hline 
		Air 3D & $30.8654 \pm 0.1351$ & $3.0881$ & $129.1165\pm 0.13$ & $12.6373$  & $134.77$ & $16.8462$ \\ \hline
		Starlings & $8.6889 \pm 0.8323$  & $0.42853$ & $15.2693 \pm 0.167$ & $7.4387$  & N/A & N/A \\ \hline
	\end{tabular}
\end{table}
Table \ref{tab:time_compute} depicts the time it takes to run the total variation diminishing Runge-Kutta scheme for the reachable  problems considered. The column \texttt{Avg. local} is the average time of running one single step of the TVD-RK scheme (\cref{sec:temporal}) while the  \texttt{ Global } column denotes the average time to compute the full TVD-RK solution to the HJ \pde. Each time query field represents an average over 20 experiments. 
We see that computation is significantly faster when our schemes are implemented on a GPU in all categories save the low-dimensional double integral plant system. We attribute this to the little amount of data points used in the overapproximated stacked levelsets. For the Air3D game and the two rockets differential game problem, the average local time for computing the solutions to the stagewise HJ \pde's sees a gain of $\sim76\%$; the global time shows a gain of $76.09\%$ over \citet{MitchellLSToolbox}'s MATLAB computational scheme. Similarly, we notice substantial computational gains for the two rockets differential game problem: $89\%$ faster global optimization time and $88.62\%$ average local computational time compared to our \text{CPU} implementations in \texttt{Numpy}. For the rockets game, compared against~\citet{MitchellLSToolbox} library, we notice a speedup of almost $92\%$ in global optimization with the \texttt{GPU} library versus an $89.32\%$ gain using our \texttt{CPU}-\texttt{NUMPY} library. Notice the exception in the \texttt{Double Integrator} experiment, however: local and global computations take a little longer compared to deployments on the~\texttt{Numpy} CPU implementation and \citet{MitchellLSToolbox}'s native ~\texttt{MATLAB}\textregistered toolbox. We attribute this to the little size of arrays of interest in this problem. The entire target set of the double integral plant exists on a two-dimensional grid whose analytic and approximate time-to-reach-the-origin computational time involves little computational gain in passing data onto the GPU. Nevertheless, we still see noticeable gains in using our CPU implementation as opposed to~\citet{MitchellLSToolbox}'s native ~\texttt{MATLAB}\textregistered toolbox. 

On a CPU, owing to efficient arrays arithmetic native to ~\citet{Numpy}'s Numpy library, the average time to compute the zero levelsets per optimization step for the \texttt{odeCFLx} functions is faster with our Numpy implementation compared against~\citet{MitchellLSToolbox} LevelSets \texttt{MATLAB\textregistered} Toolbox library computations across all experiments. The inefficiencies of \texttt{MATLAB\textregistered}'s array processing routine in the longer time to resolve stagewise BRTs and the effective time to finish the overall HJ \pde resolution per experiment manifests in all of our experimental categories. For \texttt{CPU} processing of HJ \pde's, it is reasonable, based on these presented data to expect that users would find our library far more useful for everyday computations in matters relating to the numerical resolution of HJ \pde's.

In all, there is conclusive evidence that our implementations are faster, extensible to modern libraries, and  scalable for modern complex system design and verification problems that arise. 
\textcolor{blue}{\section{Conclusion}
\label{sec:conclude}
With the increasing complex systems that are gaining utilization almost every domain of expertise including reinforcement learning, control systems, robotics, modeling, and real-time prediction in sensitive and safety-critical environments, it has become paramount to start considering backup safety filters for these generally unstable large function approximators in safety-critical environments. We have presented all the essential components of the python version in the \texttt{levelsetpy} library for numerically resolving HJ PDEs (in a reachability setting and other associated contexts) by advancing co-dimension one interfaces on Cartesian grids. We have motivated the work presented with several numerical examples to demonstrate the efficacy of our library. 
The examples we have presented in this document provide a foray into the computational aspects of ascertaining the safety (freedom from harm) of the complex systems that we are continually designing and building. This includes safety-critical analysis encompassing simple and complex multiagent systems whose safe navigation over a phase space can be considered in an HJ \pde framework. As complexity evolves, we hope that our library can serve as a  useful tool for tinkerers looking for an easy-to-use proof-of-concept toolkit for verification of dynamical systems based on a principled numerical analysis. 
We encourage users to download the library from the author's github webpage (it is available in \texttt{CPU} and \texttt{GPU} implementations via appropriately tagged branches) and use it robustly for various problems of interest where speed and scale for the solubility of hyperbolic conservation laws and HJ PDE's are of high importance. }

\bibliographystyle{acm}
\bibliography{levtoms}

\end{document}